\documentclass[twocolumn,twocolappendix,iop,apj]{CBD-simulation}

\usepackage{amsmath}

\newcommand{\m}[1]{\textcolor{purple}{#1}}

\submitjournal{APJ}

\shorttitle{Hydro Simulation of CBD}
\shortauthors{Wang et al.}

\graphicspath{{./}{figures/}}

\begin{document}

\title{On the Role of Gas Cooling in the Dynamics of Circumbinary Disks}

\author[0000-0001-7167-6110]{Hai-Yang Wang}
\affiliation{Fudan University, Department of Physics, Shanghai 200433, China; \href{cgmgalaxy0721@gmail.com}{cgmgalaxy0721@gmail.com}}

\author[0000-0001-6906-9549]{Xue-Ning Bai}
\affiliation{Institute for Advanced Study and Department of Astronomy, Tsinghua Univeristy, Beijing 100084, China; \href{xbai@tsinghua.edu.cn}{xbai@tsinghua.edu.cn}} 

\author[0000-0002-1934-6250]{Dong Lai} 
\affiliation{Department of Astronomy, Center for Astrophysics and Planetary Science, Cornell University, Ithaca, NY 14853, USA; \href{dong@astro.cornell.edu}{dong@astro.cornell.edu}}

\begin{abstract}

Hydrodynamical interactions between binaries and circumbinary disks
(CBDs) play an important role in a variety of astrophysical systems,
from young stellar binaries to supermassive black hole binaries.
Previous simulations of CBDs have mostly employed locally isothermal
equation of state. We carry out two-dimensional viscous hydrodynamic
simulations of CBDs around equal-mass, circular binaries, treating the
gas thermodynamics by thermal relaxation towards equilibrium
temperature (the constant-$\beta$ cooling ansatz, where $\beta$ is the
cooling time in units of the local Keplerian time).
As an initial study, we use the grid-based code {\texttt{Athena++}} on a polar
grid, covering an extended disk outside the binary co-orbital region.
We find that with a longer cooling time, the accretion variability is
gradually suppressed, and the morphology of the CBD becomes more
symmetric. The disk also shows evidence of hysterisis behavior depending on
the initial conditions. Gas cooling also affects the rate of
angular momentum transfer between the binary and the CBD, where given our
adopted disk thickness and viscosity ($H/r\sim 0.1$ and $\alpha\sim 0.1$),
the binary orbit expands while undergoing accretion for most $\beta$ values
between 0 and 4.0 except over a narrow range of intermediate $\beta$ values.
The validity of using polar grid excising the central domain is also discussed.
\end{abstract}


\section{Introduction} \label{sec:intro}

As a common occurrence in nature, circumbinary disks (CBDs) play an important role in a wide range of astrophysical systems. Understanding the interaction between the binary and the disk is essential in order to decipher the mystery of binary orbital evolution, the overall CBD evolution and observational properties, and physical processes that take place within CBDs \citep{2022arXiv221100028L}.

{CBDs are thought to influence the orbital evolution of binary supermassive black holes (SMBHs), which eventually determines their merger rates. In particular, it is generally known that dynamical friction by gas and stars and stellar ``loss-cone" scattering would serve to reduce the separation of binary SMBHs following galaxy mergers to of the order one parsec, while gravitational radiation would shrink binary's orbit when their distance is less than $\sim$ 0.01 pc (e.g. \citealt{2003ApJ...583..616J,2003ApJ...590..691W,2008MNRAS.390..192S,2017MNRAS.464.3131K}). 
Orbital evolution in the intermediate regime from $\sim 1$ to $\sim 0.01$ pc separations, is a long-standing problem known as the ``final parsec problem" \citep{1980Natur.287..307B}. 
It is generally believed that the process is gas-assisted, thanks to the presence of CBDs orbiting binary SMBHs (e.g. \citealt{2002ApJ...567L...9A,2008ApJ...672...83M,2009ApJ...700.1952H}).}

Circumbinary disks are also expected to form around young stellar binaries as a byproduct of binary star formation~\citep{1986ApJS...62..519B,1994MNRAS.271..999B,1994MNRAS.269L..45B,2008ApJ...681..375K}.
There are several observations on disks around Class I/II young stellar binaries, such as GG Tau, DQ Tau and UZ Tau E~\citep{1994A&A...286..149D,1996A&AS..117..393B,1997AJ....113.1841M} and around younger Class 0 stellar binaries like L1448 IRS3B \citep{2016Natur.538..483T}. {The presence of CBDs affect the binary orbital evolution, and binary leaves observable imprints on the CBDs. Perhaps more interestingly, the dynamics of such CBDs can affect the formation and migration of circumbinary planets \citep{2014A&A...564A..72K,2015A&A...581A..20K,2017MNRAS.465.4735M,2018A&A...616A..47T,2021A&A...645A..68P}.
}

{Early analytical studies found that the interaction of a binary with the outer CBD makes the binary lose angular momentum, causing orbital decay~\citep{1991MNRAS.248..754P}. Similar to the classical theory of planet-disk interaction~\citep{1979MNRAS.186..799L,1986ApJ...309..846L,1979ApJ...233..857G,1980ApJ...241..425G}, the binary and disk exchange angular momentum by coupling through the torques generated at the Lindblad resonances~\citep{1979ApJ...233..857G}. 
However, analytically estimating the angular momentum flux between the binary and the disk is challenging, mostly due to our ignorance of the nonlinear fluid motion in the close vicinity of the binary. 
Long-term numerical simulations are more suitable for studying this problem.}

There have been a large amount of simulation works studying the dynamics of CBDs, most of which are conducted in 2D viscous hydrodynamics (see, e.g. \citealt{1996ApJ...467L..77A,2002A&A...387..550G} for earlier works; more references can be found in \citealt{2019ApJ...871...84M}). 
It was recognized that long-term simulations are needed to achieve a quasi-steady state before diagnostics are applied for measuring the disk properties and determining the secular binary evolution (e.g. \citealt{2016ApJ...827...43M,2017MNRAS.466.1170M}).
The parameter space in typical CBD problem is extensive, including binary eccentricity $e$ (e.g. \citealt{2016ApJ...827...43M,2019ApJ...871...84M,2020ApJ...889..114M,2021ApJ...914L..21D,2021ApJ...909L..13Z}), mass ratio $q$ (e.g. \citealt{2020ApJ...889..114M,2020ApJ...901...25D}), disk aspect ratio $h=H/r$~\citep{2022MNRAS.513.6158D,2020A&A...641A..64H,2020ApJ...900...43T}, and the viscosity prescription~\citep{2020ApJ...901...25D}, and they are still being actively explored. 
Contrary to earlier expectations, it is now clear that the dynamics of CBDs and their various manifestations can depend significantly on these parameters.
If we focus ourselves on near-equal-mass binaries on circular orbits, it has been shown that the binary can outspiral for CBDs having disk aspect ratio $h=0.1$~\citep{2017MNRAS.466.1170M,2019ApJ...871...84M,2020ApJ...889..114M}. More recent works~\citep{2022MNRAS.513.6158D,2020A&A...641A..64H,2020ApJ...900...43T} found that on the other hand, the binary undergoes inspiral in the presence of thinner CBDs ($h\lesssim0.04$, 
more appropriate for disks around SMBHs).

While CBDs have been studied over a wide range of parameter space, disk thermodynamics is almost exclusively treated in a highly simplified manner, that is, assumed to be locally isothermal.
More specifically, the disk temperature profile is a prescribed function of position, considered to be the equilibrium temperature of the system, usually chosen to maintain a constant disk aspect ratio $h$. This treatment is convenient, but the underlying assumption is an extremely efficient cooling process that instantly brings disk temperature back to equilibrium.

Without resorting to radiative transfer, the simplest way to assess the role of thermodynamics is to allow the disk temperature to relax towards the equilibrium value over some cooling timescale $t_{\text{cool}}$. In the case of planet-disk interaction (i.e., extremely small mass ratio), finite cooling time has been found to more effectively damp the density waves, leading to changes in the deposition of angular momentum from density waves and disk morphologies
\citep{2020ApJ...892...65M,2020MNRAS.493.2287Z}.
Only a couple recent works on CBDs have included such cooling prescriptions, but these works either focused on other physics such as
self-gravity~\citep{2021MNRAS.507.1458F}, or aimed at modeling specific astrophysical systems~\citep{2018A&A...616A..47T,2019A&A...627A..91K}. \citet{2015MNRAS.446L..36F} and \citet{2018MNRAS.476.2249T} employed radiative cooling to study the thermal emission from circumbinary disks around supermassive black hole binaries. In the regime near binary merger, cooling is also included in works such as \citet{2012ApJ...755...51N} and \citet{2018ApJ...865..140D}.
The roles of thermodynamics on central topics of CBD dynamics, particularly disk morphology, accretion rate variability and binary orbital evolution, are yet to be explored. 
Towards the end of our study, we learned that \citet{2022A&A...664A.157S} thoroughly investigated the influence of cooling on thin circumbinary disk, focusing on cavity shape and size. Our work can serve as a complementary study with a different parameter space and focus.

In this work, we conduct a set of 2D viscous simulations of CBDs, aiming to examine the impact of cooling timescale on the dynamics of CBDs. As an initial study, we restrict ourselves to equal-mass and circular binaries, and fix the (equilibrium) disk aspect ratio to $h=0.1$. Our main focus is to determine how the properties of CBDs and the resulting binary evolution depend on our imposed cooling time, illustrated by a series of controlled experiments.
We note that there are two types of simulations of CBDs: the first adopts Eulerian methods in cylindrical coordinates that excise the central region (e.g. \citealt{2008ApJ...672...83M,2012ApJ...749..118S,2013MNRAS.436.2997D,2017MNRAS.466.1170M,2021ApJ...922..175N}), 
and the second uses Eulerian methods in cartesian coordinates or Lagrangian methods (such as SPH and moving mesh codes) with computation domain covering the entire region including the binary orbit (e.g. \citealt{1996ApJ...467L..77A,2009MNRAS.393.1423C,2012A&A...545A.127R,2013MNRAS.429..895P,2015MNRAS.448.3545D,2016ApJ...827...43M,2019ApJ...871...84M,2019ApJ...875...66M,2020ApJ...889..114M,2020ApJ...901...25D}). The former is computationally more efficient, and better conserves angular momentum for the outer CBDs, which we adopt in this work. We also note that this approach is valid when materials entering the co-orbital region of central binary do not get ejected across the chosen inner boundary, as has been found to be the case for the parameters that we choose~\citep{2019ApJ...871...84M,2022ApJ...932...24T}. In a companion paper \citep{paper1}, we will expand the parameter space and make our computational domain enclose the central binary.

This paper is organized as follows. In Section \ref{sec:setup}, we summarize our simulation setup and analysis procedures. In Section \ref{sec:validation}, we validate our setup with locally isothermal simulation. In Section \ref{sec:results}, we present the simulation results to illustrate the impact of disk thermodynamics. Finally, we discuss possible astrophysical implications and conclude in Section \ref{sec:conclusion}.

\section{Problem Setup} \label{sec:setup}

\subsection{Simulation Setup}
\label{subsection:simulation-setup}
We use the grid-based Godunov code \texttt{ ATHENA++} \citep{2020ApJS..249....4S} to solve the vertically integrated viscous hydrodynamic equations in cylindrical coordinates $(r,\phi)$. The main equations in conservative form read
\begin{align} 
    & \frac{\partial \Sigma}{\partial t} + \nabla \cdot (\Sigma \bf{v}) = 0 , \\ 
    & \frac{\partial (\Sigma \bf{ v})}{\partial t} + \nabla \cdot (\Sigma {\bf vv} + P \mathcal{I} + {\mathcal T}_{\rm{visc}})= -\Sigma \nabla \Phi , \\
    & \frac{\partial E}{\partial t} + \nabla\cdot[(E+P+{\mathcal T}_{\rm{visc}}){\bf v}] = -\Sigma {\bf v}\cdot\nabla\Phi 
    +\Lambda ,
\end{align}
where $\Sigma$, $\mathbf{v}$ are the disk surface density and velocity, $P$ is the vertically integrated pressure, $\mathcal{I}$ is the identity tensor, $\Phi$ is the gravitational potential, and ${\mathcal T}_{\text{visc}}$ is the viscous stress tensor given by 
\begin{equation}
    \mathcal{T}_{\mathrm{visc}, i j}=-\Sigma \nu\left(\frac{\partial v_{i}}{\partial x_{j}}+\frac{\partial v_{j}}{\partial x_{i}}-\frac{2}{3} \frac{\partial v_{k}}{\partial x_{k}} \delta_{i j}\right) ,
\end{equation}
with $\nu$ the kinematic viscosity (to be specified later). The total energy density is given by
\begin{equation}
    E=\frac{1}{2}\Sigma v^2+\frac{P}{\gamma-1}\ ,
\end{equation}
where $\gamma=5/3$ is the adiabatic index, and $\Lambda$ is the cooling term (to be specified later). Viscous heating process is automatically included in the code, entering the  energy equation in the form of (the viscous heating rate)
\begin{equation}
    Q_{\rm{visc}} = - \nabla \cdot ({\mathcal T}_{\rm{visc}} {\bf v})
\end{equation}
We assume equal-mass binary on a circular orbit,
with the mass of the two gravitating objects $M_1=M_2$, and the binary semi-major axis $a_{\rm{B}}$.
Our coordinate system centers on the center-of-mass of the binary with computational domain extending from $r_{\text{in}}=a_{\rm{B}}$ to $r_{\text{out}}=70a_{\rm{B}}$. The only exception is that we set $r_{\rm in}=0.75a_{\rm{B}}$ in one test Run T.
The gravitational potential can be expressed as 
\begin{equation} 
\Phi(r,\phi,t) = - \sum_{i=1}^{2} \frac{GM_i}{[r^2+r_i^2-2rr_i\cos(\phi-\phi_i)]^{\frac{1}{2}}}
\end{equation}
with the location of each binary component $(r_i,\phi_i)|_{i=1,2}$. The binary rotates counter-clockwise with Keplerian speed and the CBD evolves under this rotating potential.

The disk temperature is defined as 
\begin{equation}
    T\equiv P/\Sigma=c_s^2\ ,
\end{equation}
where $c_s$ is the isothermal sound speed. We set the equilibrium temperature $T_{\rm eq} = h^2r^2\Omega_{\rm{K}}^2$ such that the disk aspect ratio $h=H/r=0.1$, where $H=c_s/\Omega_{\rm{K}}$ is the disk scale height, and $\Omega_{\rm{K}}=(GM_{\rm{B}}/r^3)^{{1}/{2}}$ is the Keplerian angular velocity at $r$, with $M_{\rm{B}}=M_1+M_2$ being the total mass of the binary.
As the system evolves, we thermally relax gas temperature to $T_{\rm eq}$ over a cooling timescale $t_{\rm cool}$, controlled by the cooling term
\begin{equation} 
\Lambda=-\frac{\Sigma}{\gamma - 1} \times \frac{(T - T_{\text{eq}})}{t_{\text{cool}}}\ .
\label{eq:cooling}
\end{equation}
We prescribe the cooling time in a scale-free manner, using the dimensionless parameter
\begin{equation}
    \beta\equiv\Omega_{\rm{K}}t_{\rm cool}\ ,
\end{equation}
known as $\beta-$cooling \citep{2001ApJ...553..174G}. 

With this temperature prescription, we set the kinematic viscosity according to the standard $\alpha$ prescription \citep{1973A&A....24..337S}
\begin{equation} 
\nu = \alpha H c_s\ .
\end{equation}
We fix $\alpha=0.1$ for all simulations. We emphasize that by default, as gas temperature varies in our simulations, viscosity varies accordingly (see Section \ref{subsection:models}). 
Note that in the steady-state (and neglecting the perturbation from the binary tidal potential), the viscous heating rate can be approximated by $Q_{\rm{visc}}=(9/4)\nu \Sigma \Omega_{\rm{K}}^2$. 
Energy balance in the CBD requires
\begin{equation}
    Q_{\rm{visc}}+\Lambda = 0 ,
\end{equation}
This implies that the steady-state disk temperature $T_{\rm{st}}$ is given by (see also \citealt{2022A&A...664A.157S})
\begin{equation}
1- \frac{T_{\rm{eq}}}{T_{\rm{st}}}= \frac{9}{4} (\gamma-1)\alpha\beta
\label{eq:teq}
\end{equation}
When choosing $\alpha=0.1$ and $\gamma=5/3$, the condition that $T_{\rm{st}}$ is finite (or $1- T_{\rm{eq}}/T_{\rm{st}} <1$) implies $\beta \lesssim 6$. In this study, we survey the range $\beta \in [0.0, 4.0]$. When slightly deviating from locally isothermal condition (which corresponds to $\beta=0$), 
steady-state temperature $T_{\rm{st}}$ is somewhat larger than $T_{\rm{eq}}$.

We choose the initial condition similar to \citet{2019ApJ...871...84M}, with disk surface density given by 
\begin{equation} 
\begin{aligned}
    \Sigma(r) = &\frac{\dot M_0}{3\pi\alpha h^2 \sqrt{GM_{\rm{B}}r}} \left[1 - \frac{l_0}{\Omega_{\rm{B}}a_{\rm{B}}^2}\sqrt{\frac{a_{\rm{B}}}{r}}\right] \\
    &\rm{exp}\left[-\left(\frac{r}{r_{\text{edge}}}\right)^{-12}\right]\ ,
\end{aligned}
\end{equation}
where $l_0=0.7\Omega_{\rm{B}} a_{\rm{B}}^2$ is the approximate value of the specific angular momentum, $r_{\rm edge}\equiv2r_{\rm in}\equiv2a_{\rm{B}}$ {is the anticipated radial location of cavity edge}. The initial rotation profile of the disk is assumed to be Keplerian (over a single center mass of $M_{\rm{B}}$), and the initial radial velocity is set to
\begin{equation} 
    v_r(r) = - \frac{3\nu}{2r} \left[1-\left(\frac{r_{\text{in}}}{r}\right)^{\frac{1}{2}}\right]^{-1}\ ,
\end{equation}
which is the expected accretion velocity in a steady-state viscously-driven disk with a single central object.

At the outer radial boundary, we use a fixed state boundary condition as in the initial condition, which ensures a steady mass injection rate $\dot M_0$.
At the inner boundary, we use the ``diode" boundary condition as in \citet{2017MNRAS.466.1170M}, where zero gradient condition is employed on surface density $\Sigma$, radial velocity $v_r$, and azimuthal velocity $v_{\phi}$ by default, but if $v_r$ is positive, its value is set to zero so that mass can only exit the inner boundary, while mass entering from the inner boundary is prohibited.

Our simulations are scale-free. In code units, we choose $GM_{\rm{B}}=a_{\rm{B}}=\Omega_{\rm{B}}=1$, and $\dot M_0 = 3\pi\alpha h^2$. 
{Consequently, the binary orbital period is $P_{\rm{B}}=2\pi /\Omega_{\rm{B}}=2\pi$.}
We use logarithmic grid in the radial direction and uniform grid in the azimuthal direction,
keeping cell aspect ratio $\Delta r \approx 0.8 r \Delta \theta$. For all simulations except one, we have $N_{r}\times N_{\phi} = 432 \times 504$ cells, so that the resolution is 8 cells per scale height, and the grid size along the radial direction at $r=a_{\rm{B}}$ is $\Delta r \simeq 0.010 a_{\rm{B}}$.
For numerical stability, we employ a density floor $\Sigma_{\text{min}} = \text{min} (10^{-6}\Sigma(r), \Sigma_{\text{min}}^{\text{const}})$, where $\Sigma_{\text{min}}^{\text{const}} = 10^{-4}$ in simulation units. We also adopt a temperature floor {$T_{\text{min}} = 10^{-3} T_{\text{eq}}$}, along with a temperature cap assuming disk aspect ratio $h$ cannot exceed 0.5 throughout simulation.

By employing a cylindrical grid, our simulation ensures angular momentum conservation to high accuracy, while at the cost of excising a hole in the center of the domain. 
We note that while many recent simulations of CBDs fully enclose the binary in the simulation domain, the properties of the CBDs, particularly the angular momentum flux through the disk, is insensitive to this choice as long as there is no significant mass ejection from the central binary. Our choice of parameters are not subject to such phenomenon, as justified in \citet{2022ApJ...932...24T}, as well as by comparing simulation results of \citet{2017MNRAS.466.1170M} and \citet{2019ApJ...871...84M} for the same parameters.

\subsection{Simulation Runs}
\label{subsection:models}

As an initial survey of the parameter space, we carry out a series of simulations with a fixed $\beta$. The main simulation parameters are summarized in Table \ref{tab:simulation-parameters}. To more thoroughly explore the impact of dynamical cooling, we run a series of long-term simulations with a slowly varying $\beta$ listed in Table \ref{tab:simulation-parameters-longterm}.

Focusing on the fixed-$\beta$ runs in Table \ref{tab:simulation-parameters}, among all simulation runs, F0, T0 and T are listed for the purpose of validation. Run F0 reproduced the results adopting locally isothermal condition similar to \citet{2017MNRAS.466.1170M}. Run T0 is identical to F0 except using low resolution $N_r \times N_{\phi} = 160 \times 210$. Run T extends the radial location of the simulation inner boundary to $r_{\text{in}} = 0.75 a_{\rm{B}}$, with $\beta=1.0$.

As we will discuss in Section \ref{section:hysteresis}, we found that the results of binary evolution may depend on whether the density profile of the disk is axisymmetric or not when the simulation starts. We thus separately run simulations having the same cooling time (with constant $\beta$), but with non-axisymmetric and axisymmetric initial conditions. Run  F0.8, F0.9, F1.0 are restarted with the quasi-steady profile (non-axisymmetric) of Run F0. 
Simulations F0, B0.1, B0.6, B0.7, B0.8, B1.0 are started with the initial condition described in Section \ref{subsection:simulation-setup}. 

Among the long-term simulations listed in Table \ref{tab:simulation-parameters-longterm}, we first focus on two fiducial runs. In Run F and Run B, we slowly change $\beta$ from 0 to 1.3 and from 1.3 to 0. We first run for 2,000 binary orbits at a fixed $\beta$ (either $0$ or $1.3$) to reach quasi-steady state. We then linearly vary $\beta$, which changes by 0.1 in every 2,000 binary orbits. Various torque components in the CBD are measured through out the simulations. In simulation B, to smooth the transition in angular momentum transfer profile (which we shall see in Fig \ref{fig:am-sum}), we double the timescale of changing $\beta$, namely, $\beta$ changes 0.1 every 4,000 binary orbits. 

Other long-term simulations listed in Table \ref{tab:simulation-parameters-longterm} are carried out in order to check the sweep speed when $\beta \in [0.6,1.0]$ and to probe the parameter space $\beta \in [1.0,4.0]$. These simulations are restarted from Run B1.0. In Run B-Q, B-M/B, and B-S (which we refer to as `B-series' for short), we slowly and linearly change $\beta$. When $\beta$ decreases 0.1 ($\Delta \beta = 0.1$), the simulation time are separately $2000, 4000, 6000 P_{\rm{B}}$. In Run F-Q, F-M, and F-S (`F-series'), we increase $\beta$ from 1.0 to 4.0 according to the following scheme
\begin{equation}
    \log_{10} {\beta} = \frac{t}{10\times t_{\rm{interval}}}
\end{equation}
where $t_{\rm{interval}}$ for different runs are separately denoted in Table \ref{tab:simulation-parameters-longterm}. The sweep speeds of the B-series at $\beta=1.0$ are the same as those of the F-series. This parameter sweeping scheme is referred to as `Log' in the last column of Table \ref{tab:simulation-parameters-longterm}.


\subsection{Diagnostics}
\label{subsection:analysis}

Our analysis of the simulations focus on the torque balance as the main diagnostics of angular momentum transport in the system.
The advective, viscous, gravitational and total torques along with the accretion rate are given by \citep{2017MNRAS.466.1170M,2019ApJ...875...66M}:
\begin{flalign}
\vspace{-0.5cm}
{\boldsymbol{\dot J}}_{\mathrm{adv}} &= \oint \Sigma \boldsymbol{v} \times \boldsymbol{r}(\boldsymbol{v} \cdot \boldsymbol{r}) d\phi \\
&= - \boldsymbol{\hat e_z} \oint r^2 \Sigma v_r v_{\phi} d\phi \nonumber \\
{\boldsymbol{\dot J}}_{\mathrm{visc}} &= -\oint\left(\mathcal{T}_{\mathrm{visc}} \cdot {\boldsymbol{r}}\right) \times \boldsymbol{r} d\phi \\
&= - \boldsymbol{\hat e_z} \oint r^3 \nu \Sigma \left[\frac{\partial}{\partial r}\left(\frac{v_{\phi}}{r}\right) + \frac{1}{r^2}\frac{\partial v_r}{\partial \phi}\right] d\phi \nonumber \\
\boldsymbol{T}_{\mathrm{grav}} &=\int_{r}^{r_{\rm{out}}} \left[- \oint \Sigma(\boldsymbol{r} \times \boldsymbol{\nabla} \Phi) d\phi \right] dr\\
&= \boldsymbol{\hat e_z} \int_r^{r_{\mathrm{out}}} \left(- \oint r \Sigma \frac{\partial \Phi}{\partial \phi} d\phi\right) dr \nonumber \\
{\boldsymbol{\dot J}}_{\mathrm{tot}} &= {\boldsymbol{\dot J}}_{\mathrm{adv}}-{\boldsymbol{\dot J}}_{\mathrm{visc}}-\boldsymbol{T}_{\mathrm{grav}} \\
\dot{M} 
&= -\oint \Sigma v_r r d\phi
\vspace{-0.5cm}
\end{flalign}
These torques are calculated in each timestep, and are averaged over run-time during the entire simulation. Except for the long-term simulations, we use data from the last 250 binary orbits when the inner disk has reached a quasi-steady state. In the long-term runs shown in Table \ref{tab:simulation-parameters-longterm}, the torques are averaged over 30 binary orbits.

We use time-averaged torques to determine the orbital evolution of the binary. Following the standard description of binary evolution \citep{2017MNRAS.466.1170M}
\begin{equation} 
\frac{\dot{a}}{a}=8\left(\frac{l_{0}}{l_{\rm{B}}}-\frac{3}{8}\right) \frac{\dot{M}}{M}\ ,
\end{equation}
where the specific angular momentum of the circular binary is denoted by $l_{\rm{B}}=[GM_{\rm{B}}a_{\rm{B}}]^{1/2}$, and calculated total torque per unit of the accreting mass from the simulation is $l_0 = \langle \dot J_{\mathrm{tot}} \rangle / \langle \dot M \rangle$. 
Like $\dot J_{\mathrm{tot}}$, the accretion rate $\dot M$ is also time-averaged.
The criterion for binary expansion or contraction is fully determined by whether $l_0$ is greater or smaller than $3l_{\rm{B}}/8$.

The viscous timescale at a given radius is given by $t_{\nu} = (4/9)({r^2}/{\nu})$. Over 2000-3000 binary orbits from our simulations, we anticipate that the disk should be viscously relaxed within a radius of $r_{\text{rel}} \gtrsim 9-12 a_{\rm{B}}$.
\begin{table}[htbp]
	\centering
	\begin{tabular}{cccccc}
            \hline
            \hline
            label  & $\beta$    & $N_r \times N_{\phi}$   & $r_{\text{in}}/a_{\rm{B}}$ & $t_{\rm{total}}/P_{\rm{B}}$ &\\
            \hline
            T0     & $\sim$0    & $160 \times 210$        & 1.0   & 3000 & \\
            T      & 1.0        & $432 \times 504$        & 0.75  & 3000 & \\
            \hline                        
            F0     & $\sim$0  & $432 \times 504$        & 1.0   & 3000 & \\
            F0.8   & 0.8      & $432 \times 504$        & 1.0   & 3000 & \\
            F0.9   & 0.9      & $432 \times 504$        & 1.0   & 3000 & \\
            F1.0   & 1.0      & $432 \times 504$        & 1.0   & 3000 & \\
            \hline
            B0.1   & 0.1     & $432 \times 504$        & 1.0   & 3000 & \\
            B0.6   & 0.6     & $432 \times 504$        & 1.0   & 3000 & \\
            B0.7   & 0.7     & $432 \times 504$        & 1.0   & 3000 & \\
            B0.8   & 0.8     & $432 \times 504$        & 1.0   & 3000 & \\
            B1.0   & 1.0     & $432 \times 504$        & 1.0   & 3000 & \\
            \hline
	\end{tabular}
	\caption{Simulation Parameters of Runs adopting constant cooling parameter $\beta$. The first column lists the labels to distinguish different runs. The next four columns indicate the cooling parameter $\beta$, the grid resolution in $N_r \times N_{\phi}$, the radial location of the inner boundary $r_{\rm{in}}/a_{\rm{B}}$, and the total simulation time in $P_{\rm{B}}$. }
	\label{tab:simulation-parameters}
\end{table}

\begin{table*}[htbp]
	\centering
	\begin{tabular}{ccccccc}
		\hline
		\hline
		label  & $\beta$ &$N_r \times N_{\phi}$   & $r_{\text{in}}/a_{\rm{B}}$  & $t_{\rm{interval}}/P_{\rm{B}}$ ($\Delta \beta=0.1$) & Sweep Speed Scale & \\
            \hline 
            fiducial run \\
            \hline
            F      & 0$\rightarrow$1.3   &$432 \times 504$        & 1.0                 & 2000  & Linear & \\
            B/B-M  & 1.3$\rightarrow$0.0 &$432 \times 504$        & 1.0                 & 4000 & Linear & \\
            \hline
            \hline
            F-Q    & 1.0$\rightarrow$4.0 &$432 \times 504$        & 1.0                 & 2000  & Log    & \\
            F-M    & 1.0$\rightarrow$4.0 &$432 \times 504$        & 1.0                 & 4000  & Log    & \\
            F-S    & 1.0$\rightarrow$4.0 &$432 \times 504$        & 1.0                 & 6000  & Log    & \\
		\hline
            B-Q    & 1.0$\rightarrow$0.6 &$432 \times 504$        & 1.0                 & 2000 & Linear & \\
            B-S    & 1.0$\rightarrow$0.6 &$432 \times 504$        & 1.0                 & 6000 & Linear & \\
            \hline
	\end{tabular}
	\caption{Simulation Parameters of Runs with a varying cooling parameter $\beta$, most of the which are same as Table \ref{tab:simulation-parameters}. The two additional columns  separately denote the simulation time $t_{\rm{interval}}$ when $\beta$ changes 0.1, and the scale of the sweeping speed with respect to $\beta$. Run F and B are two fiducial long-term simulations to be discussed in detail in Section \ref{sec:results}.}
	\label{tab:simulation-parameters-longterm}
\end{table*}

\section{Validation: Locally Isothermal Simulations} \label{sec:validation}

\begin{figure}[htp]
\begin{center}
\includegraphics[width=0.48\textwidth,trim={0cm 0cm 0cm .0cm},clip]{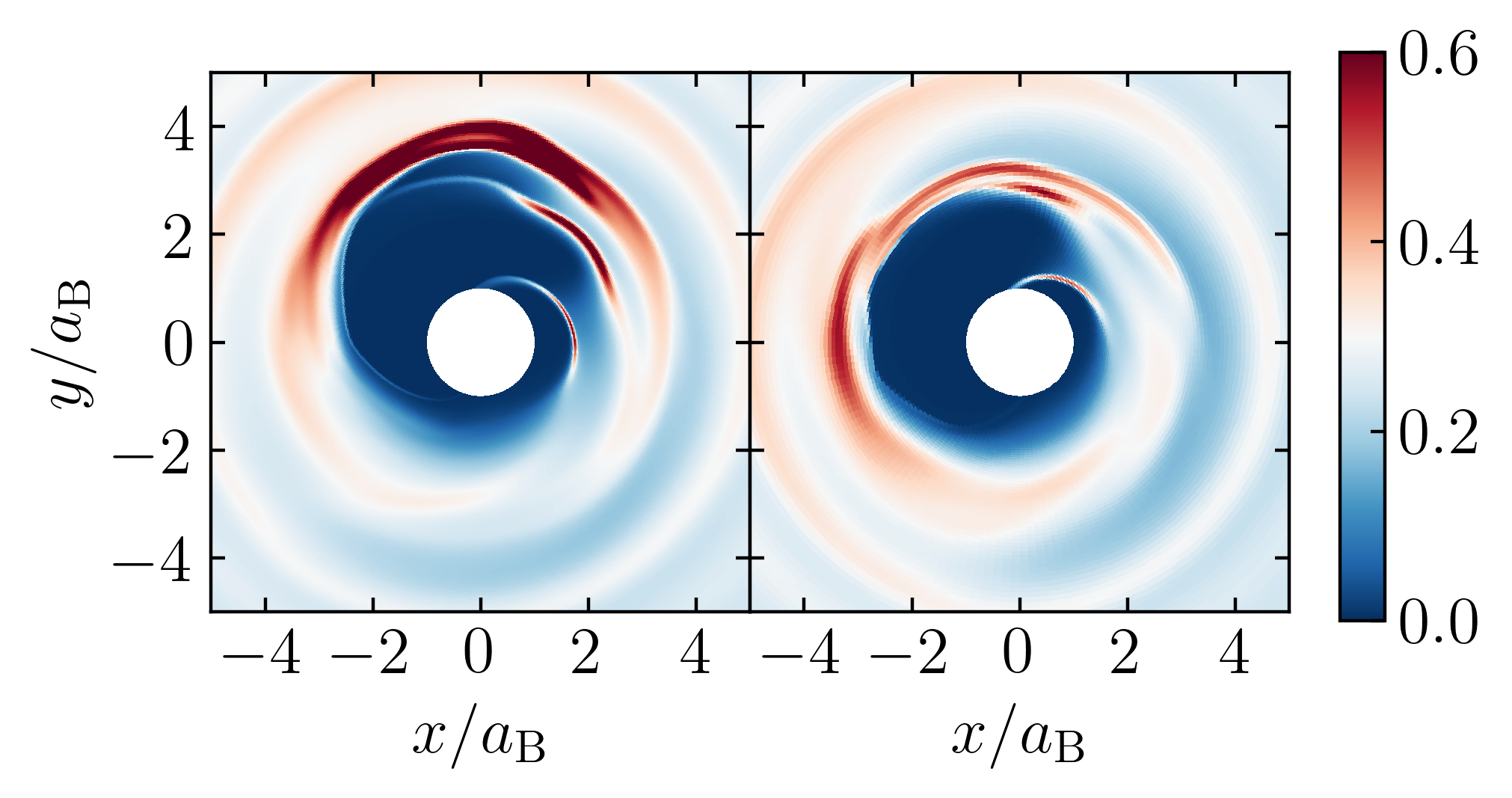}
\end{center}
\vspace{-0.3cm}
\caption{Snapshots of disk surface density from simulation F0 (left, standard resolution) and T0 (right, with a lower resolution $N_{r} \times N_{\phi} = 160 \times 210$). The morphology of the inner disk from the two runs are nearly identical. \label{fig:morphology-snap}}
\vspace{-0.3cm}
\end{figure}

{We briefly discuss the results of our fiducial simulation, employing the widely-adopted locally isothermal approximation. Our simulation setup is identical to \citet{2017MNRAS.466.1170M}, where the central region is excised. Given the physical parameters, the overall results in disk morphology, accretion variability, and angular momentum transfer are known to be consistent with simulations whose domain enclose the binary orbits and resolve the mini-disks \citep{2019ApJ...871...84M,2019ApJ...875...66M}. We analyze the accretion variability at the inner boundary and the rate of angular momentum transfer from the disk to the binary. Our analyses are conducted between 3,000 orbits and 3,250 orbits after the inner region of the disk (within $\sim 10 a_{\rm{B}}$)  has reached a quasi-steady state.}

In Figure \ref{fig:morphology-snap}, we show one characteristic snapshot of our fiducial isothermal runs F0 and T0.
The binary produces a low-density eccentric cavity with radius $\sim 2a_{\rm{B}}-3a_{\rm{B}}$, which is penetrated by narrow accretion streams feeding mass to the inner boundary. The accretion streams and the cavity rotate with the binary, nearly repeating the same pattern every $0.5 P_{\rm{B}}$.
A high-density lump near the cavity wall orbiting the binary could be easily identified.
The lump experiences periodic creation and destruction with a periodicity equal to the local Keplerian period $\sim 5P_{\rm{B}}$. All these features well reproduce the results of \citet{2017MNRAS.466.1170M}, and we defer the discussion on accretion variability, which is closely related to periodic changes in disk morphology, to Section \ref{section:accretion-variability}.

\begin{figure}[htbp]
\begin{center}
\includegraphics[width=0.48\textwidth,trim={0cm 0cm 0cm .0cm},clip]{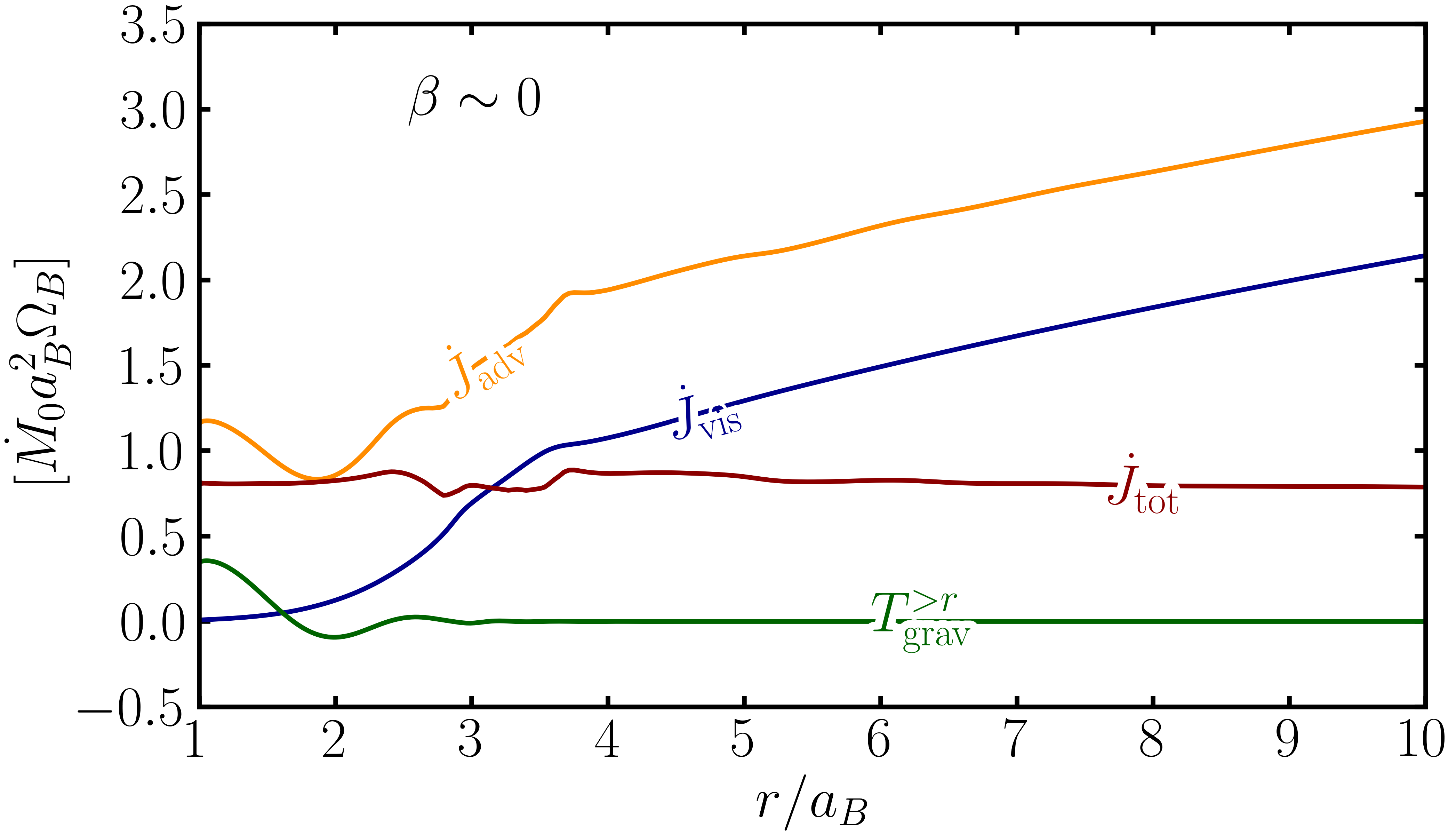}
\end{center}
\vspace{-0.3cm}
\caption{Various torque components calculated in the disk from Run F0 (locally isothermal disk), the total torque is almost flat across the radial range shown in the figure. \label{fig:am-fiducial}}
\vspace{-0.3cm}
\end{figure}

\begin{figure*}[htpb]
\begin{center}
\includegraphics[width=0.8\textwidth,trim={0cm 0cm 0cm .0cm},clip]{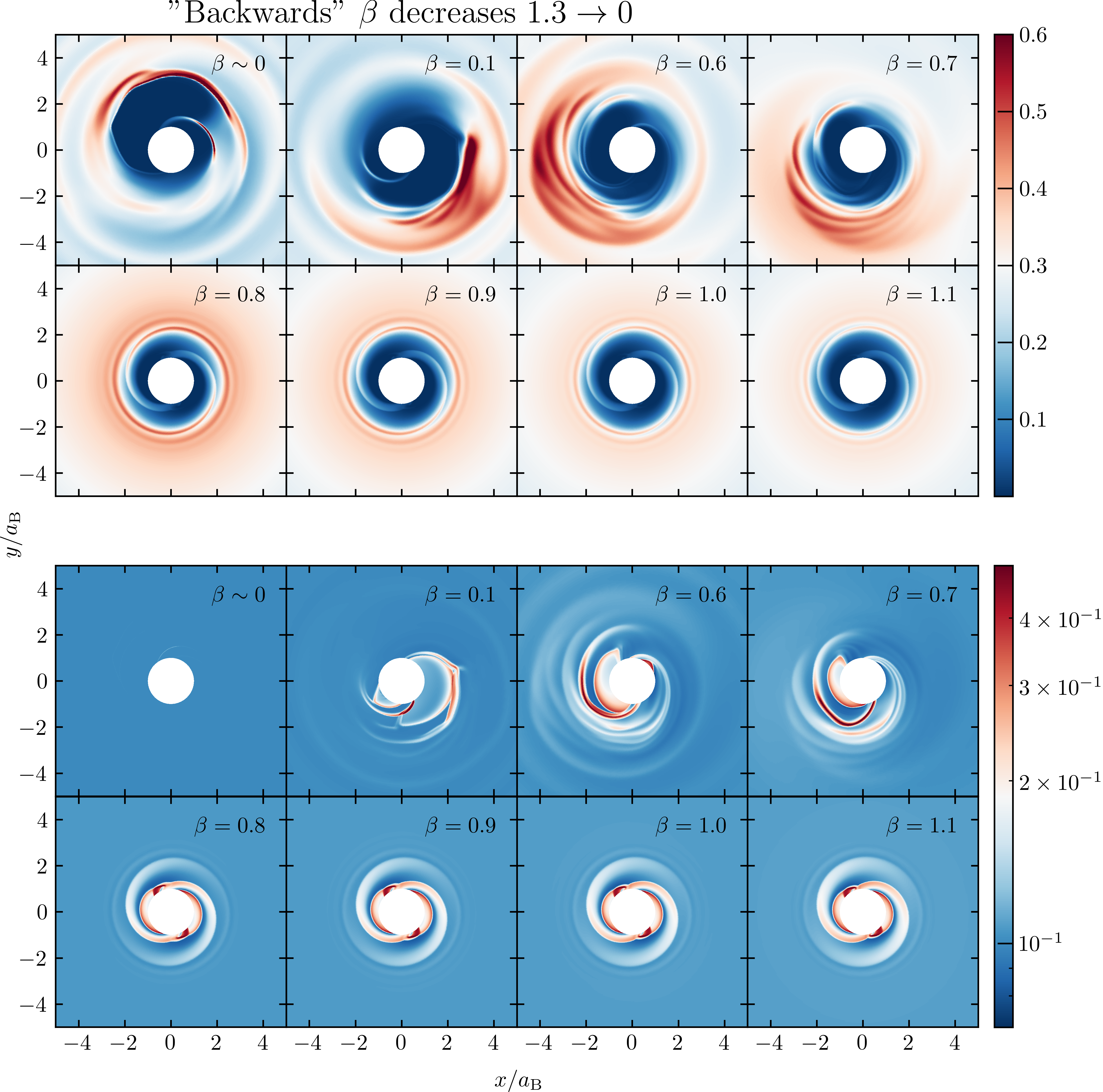}
\end{center}
\vspace{-0.3cm}
\caption{Snapshots of surface density (upper panels) and disk aspect ratio $h$ (lower panels) near the inner boundary from simulation B. They are taken as the relaxation time $t_{\mathrm{cool}}$ is slowly changed during quasi-steady state. Two transitions in disk morphology can be identified: (1) at the edge of the cavity, the lump is gradually replaced by traces of accretion streams similar to a single-armed spiral, as shown in the subfigure of $\beta = 0.6$. (2) the shape of the cavity changes from eccentric to circular with a smaller size when $\beta \gtrsim 0.7-0.8$. 
\label{fig:morphology-backward}}
\end{figure*}

\vspace{0.3cm}

We examine various torques across the CBDs from Run F0: the advective, viscous and gravitational torques, along with total torque are shown in Figure \ref{fig:am-fiducial}. The total torque is remarkably flat across the shown radial range, demonstrating that a quasi-steady state is achieved in this region. The averaged accretion rate is $\langle \dot M \rangle \simeq 1.1 \dot M_0$. The specific angular momentum carried by the accretion stream $l_0 = \langle \dot J \rangle / \langle \dot M \rangle$, is $0.86\sqrt{GM a_{\rm{B}}}$ from our simulation, in close agreement with the result of $0.82\sqrt{GM a_{\rm{B}}}$ in \citet{2017MNRAS.466.1170M}. We note that the specific angular momentum calculated in our simulation is about $25\%$ larger than the value in \citet{2019ApJ...871...84M}, most likely because of the torque exerted at the inner boundary could depend on the radial gradients which cannot be known a priori when excising the central simulation domain.\footnote{In the KITP code comparison project of binary disk interaction (Duffell et al., in preparation), it is found that the gravitational torque outside $r=a_{\rm{B}}$ from the excised simulations converges to the results from the simulations that resolve the entire domain. It is most likely that the uncertainty of the total torque calculation comes from the torque component associated with accretion onto the inner boundary.}
A snapshot of resolution test T0 is also shown in Figure \ref{fig:morphology-snap} (right panel). The central cavity, spiral density waves along with the over-dense lump are all presented, and are nearly identical to those found in Run F0. The specific angular momentum measured is also the same as that in the standard simulation.

\section{The Impact of Disk Thermodynamics} \label{sec:results}

In this section, we explore the impact of disk thermodynamics on the accretion behavior of a circular, equal-mass binary. Our discussion will be mainly based on two fiducial long-term runs B and F. First, we discuss the properties of the CBD's inner region for different $\beta$ in Section \ref{sec:result-morphology}. Then, we show the results of binary accretion variability and the binary's orbital evolution in Section \ref{section:accretion-variability} and Section \ref{section:angular-momentum-transfer}, respectively. Finally, we briefly discuss an intriguing hysteresis phenomenon for the CBD evolution in Section \ref{section:hysteresis}.

\subsection{Disk Morphologies} \label{sec:result-morphology}


Figure \ref{fig:morphology-backward} (and Figure \ref{fig:morphology}) shows a series of snapshots of surface density and disk aspect ratio $h$ from our long-term Run B (and F). 
Similar to the locally isothermal run in Section \ref{sec:validation}, a low-density cavity is cleaned up due to the presence of binary tidal torque. 
Accretion streams characterized by the primary azimuthal number $m = 1$ or $2$ feed materials to the binary. Materials piling up around the cavity wall shown in most of the subfigures are modulating accretion variability onto the binary. Details of the dependence of accretion variability on disk morphology will be discussed in Section \ref{section:accretion-variability}.


As shown in Figure \ref{fig:morphology-backward}, the increase in $\beta$ from $\sim 0$ to $1.3$ leads to two transitions in the disk morphology: In the first transition, the characteristic density distribution at the edge of the cavity wall changes from a high-density lump to a series of spirals when $\beta \sim 0.6$. The feature after this transition (i.e. for $\beta \gtrsim 0.6$) can be regarded as ``twists". If using a set of ellipses to fit the gas eccentricity (e.g. Figure 7 in \citealt{2020ApJ...905..106M}), the periapses of the ellipses are roughly in the same direction in the isothermal limit. In the case of $\beta \sim 0.6$, the periapses angle varies as the radial distance increases. These ``twists" form due to our choice of a different equation of state, and an analytical explanation will be discussed in our follow-up work. 

In the second transition, the shape of the cavity changes from eccentric to circular between $\beta \sim 0.7-0.8$. The materials piling up at the cavity wall are indicated by two-armed, tightly-wound spirals, rotating with the same angular frequency as the binary. From the lower panel of Figure \ref{fig:morphology-backward}, it is clear that the effective disk aspect ratio inside the cavity is around $0.2$, and decreases towards the prescribed value of $0.1$ at larger distances from the binary. The disk aspect ratio profile exhibits a similar morphology as the surface density profile throughout the entire $\beta$ range. As the inner cavity gets heated up, the temperature distribution also shows the feature of two-armed spirals. We note that previous simulations with a disk aspect ratio of around $0.2$ also found a similar density pattern (e.g. \citealt{2020A&A...641A..64H}), with a slightly larger pitch angle of the spirals. The circular shape of the CBD cavity indicates the absence of eccentric mode (e.g.  \citealt{2020ApJ...905..106M}) in the disk, the cause of which we will discuss in detail in our follow-up study. 

\begin{figure}[htp]
\begin{center}
\includegraphics[width=0.49\textwidth,trim={0cm 0cm 0cm .0cm},clip]{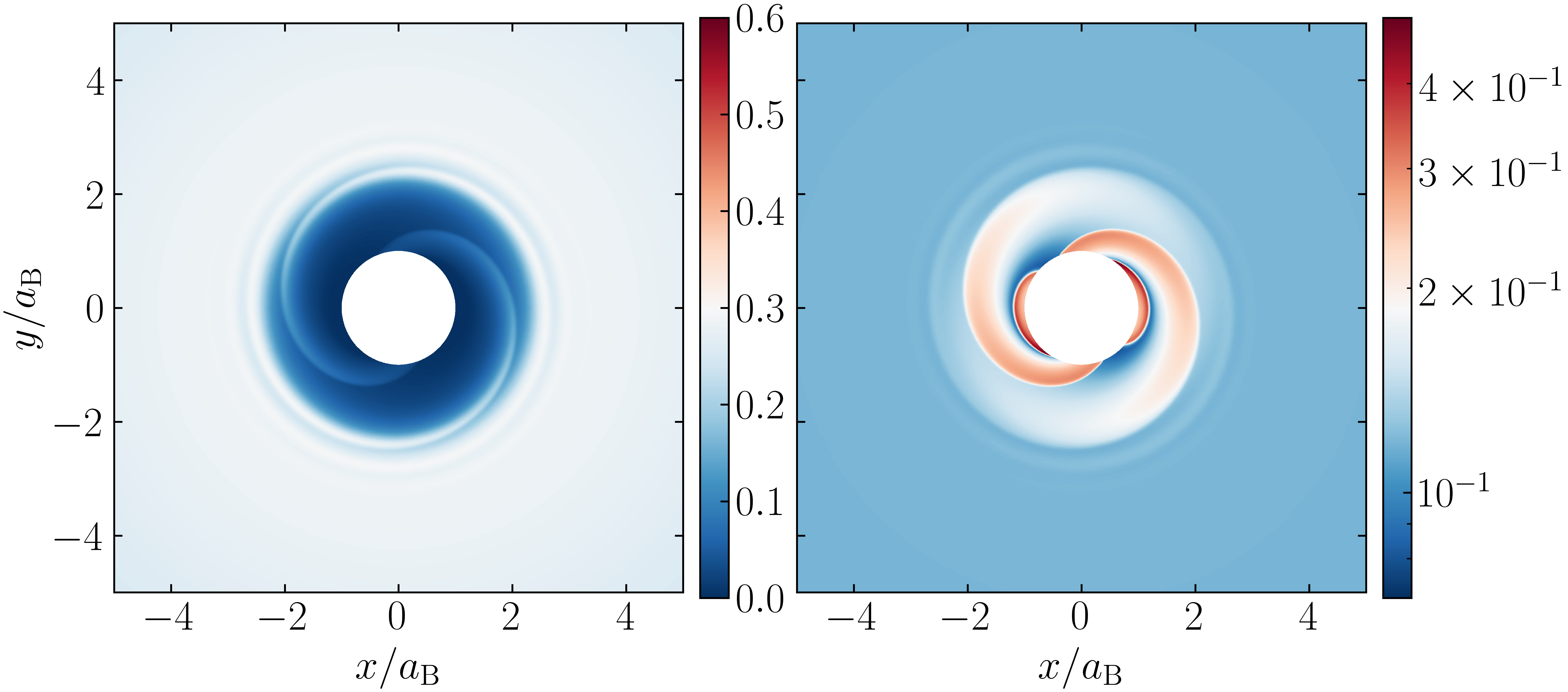}
\end{center}
\vspace{-0.3cm}
\caption{Snapshot of disk surface density (right panel) and disk aspect ratio (left panel) from long-term run F-S when $\beta=2.0$. The surface density of the inner disk remains the same as the case when $\beta \sim 1.0$, characterized by a circular cavity and two-armed, tightly-wound spirals. The disk aspect ratio profile is morphologically similar to the surface density profile. \label{fig:morphology-beta=2.0}}
\vspace{-0.3cm}
\end{figure}

By examining the F-series of runs, we find that the morphologies of the CBD are similar when $\beta \geq 1.3$. As a representative example, we show in Figure \ref{fig:morphology-beta=2.0} the surface density profile (left) and disk aspect ratio profile (right) when $\beta=2.0$. Comparing to the case of $\beta = 1.1$ in Figure \ref{fig:morphology-backward}, we see that except for the difference in the overall density and temperature scales, the spiral patterns in both density and disk aspect ratio in the two runs closely match each other.

\subsection{Accretion Variabilties} 
\label{section:accretion-variability}

\begin{figure}[htbp]
\begin{center}
\includegraphics[width=0.48\textwidth,trim={0cm 0cm 0cm .0cm},clip]{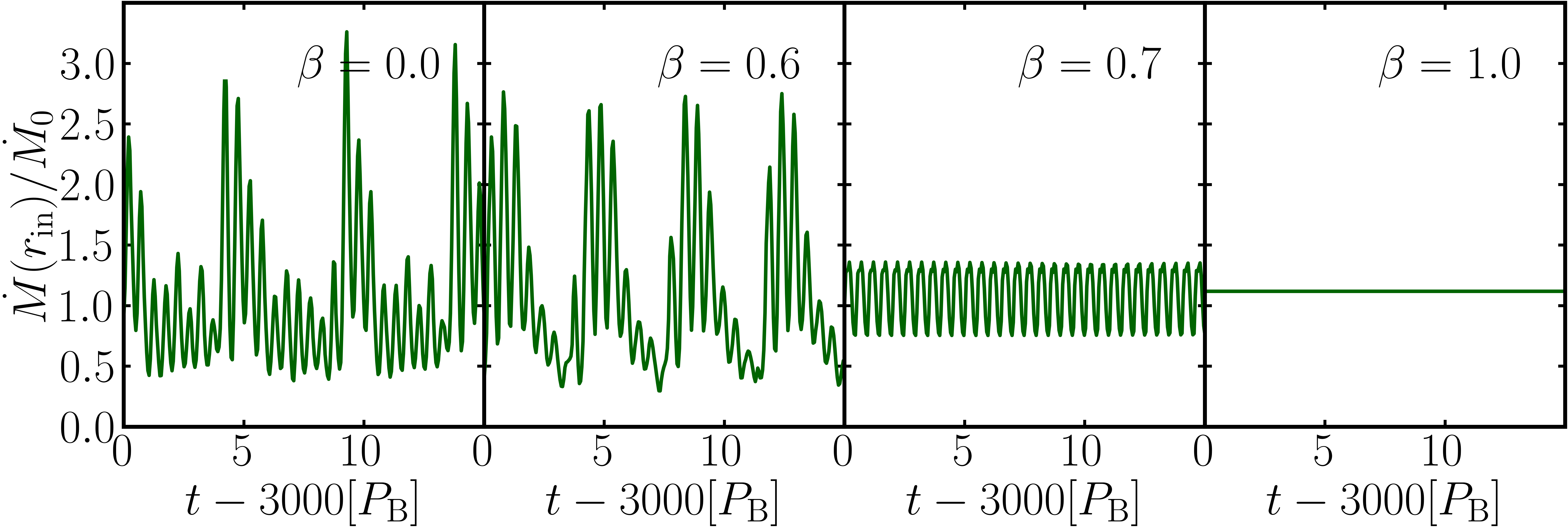}
\end{center}
\vspace{-0.3cm}
\caption{Mass accretion rate as a function of time measured at the inner boundary, for different values of $\beta=({0, 0.6, 0.7, 1.0})$. 
\label{fig:acc-vari}}
\vspace{-0.3cm}
\end{figure}

\begin{figure}[htbp]
\begin{center}
\includegraphics[width=0.4\textwidth,trim={0cm 0cm 0cm .0cm},clip]{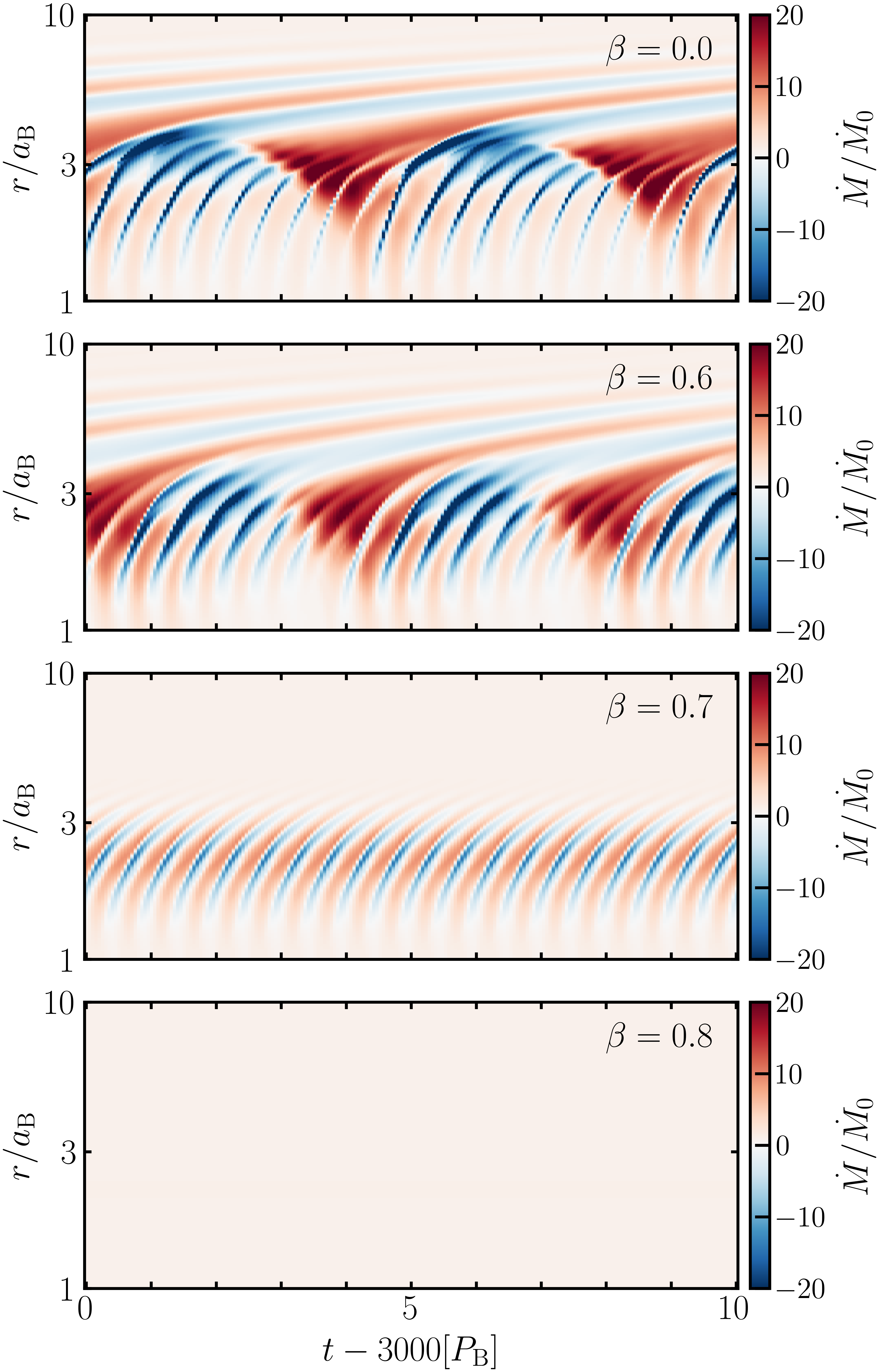}
\end{center}
\vspace{-0.3cm}
\caption{Mass accretion rate as a function of time and radius, for different $\beta$. There are two prominent transitions in accretion variability, corresponding to the suppression of the $\Omega_{\rm{B}}/5$ variability at $0.6\lesssim\beta\lesssim0.7$, and the suppression of $2\Omega_{\rm{B}}$ variability at $0.7\lesssim\beta\lesssim0.8$. \label{fig:acc-colormap}}
\vspace{-0.3cm}
\end{figure}

\begin{figure*}[htbp]
\begin{center}
\includegraphics[width=0.9\textwidth,trim={0cm 0cm 0cm .0cm},clip]{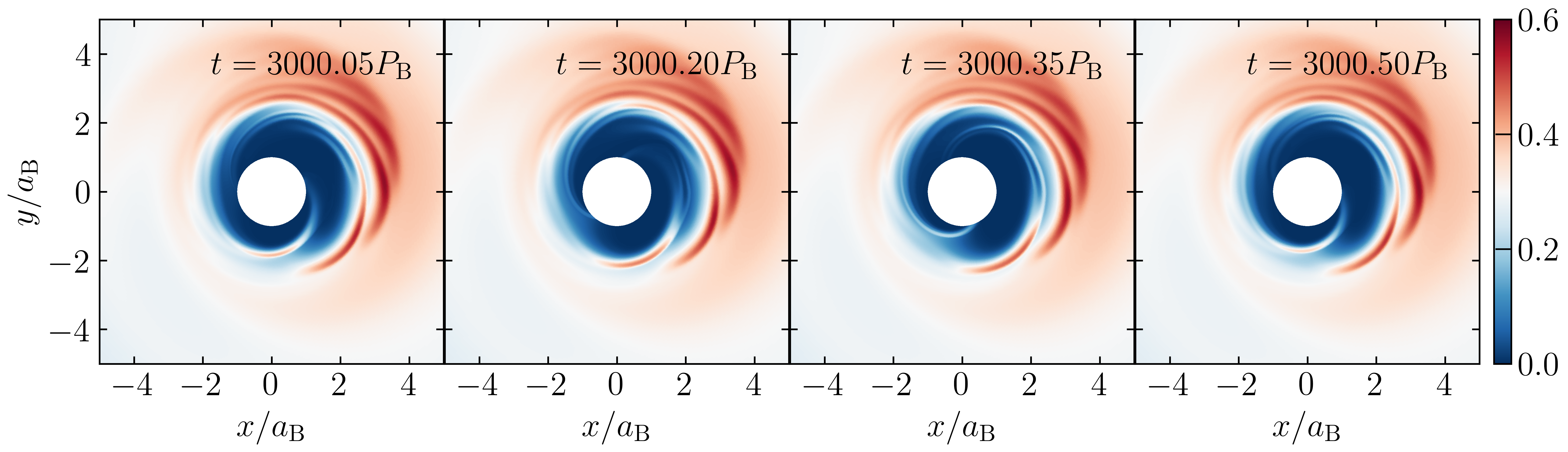}
\end{center}
\vspace{-0.5cm}
\caption{A series of surface density snapshots of simulation B0.7. The evolution process of accretion streams inside the cavity is clearly illustrated. The $2\Omega_{\rm{B}}$ accretion periodicity is directly related to this periodic morphology change.
}
\label{fig:morphology-halforbit}
\end{figure*}

\begin{figure}[htbp]
\begin{center}
\includegraphics[width=0.48\textwidth,trim={0cm 0cm 0cm .0cm},clip]{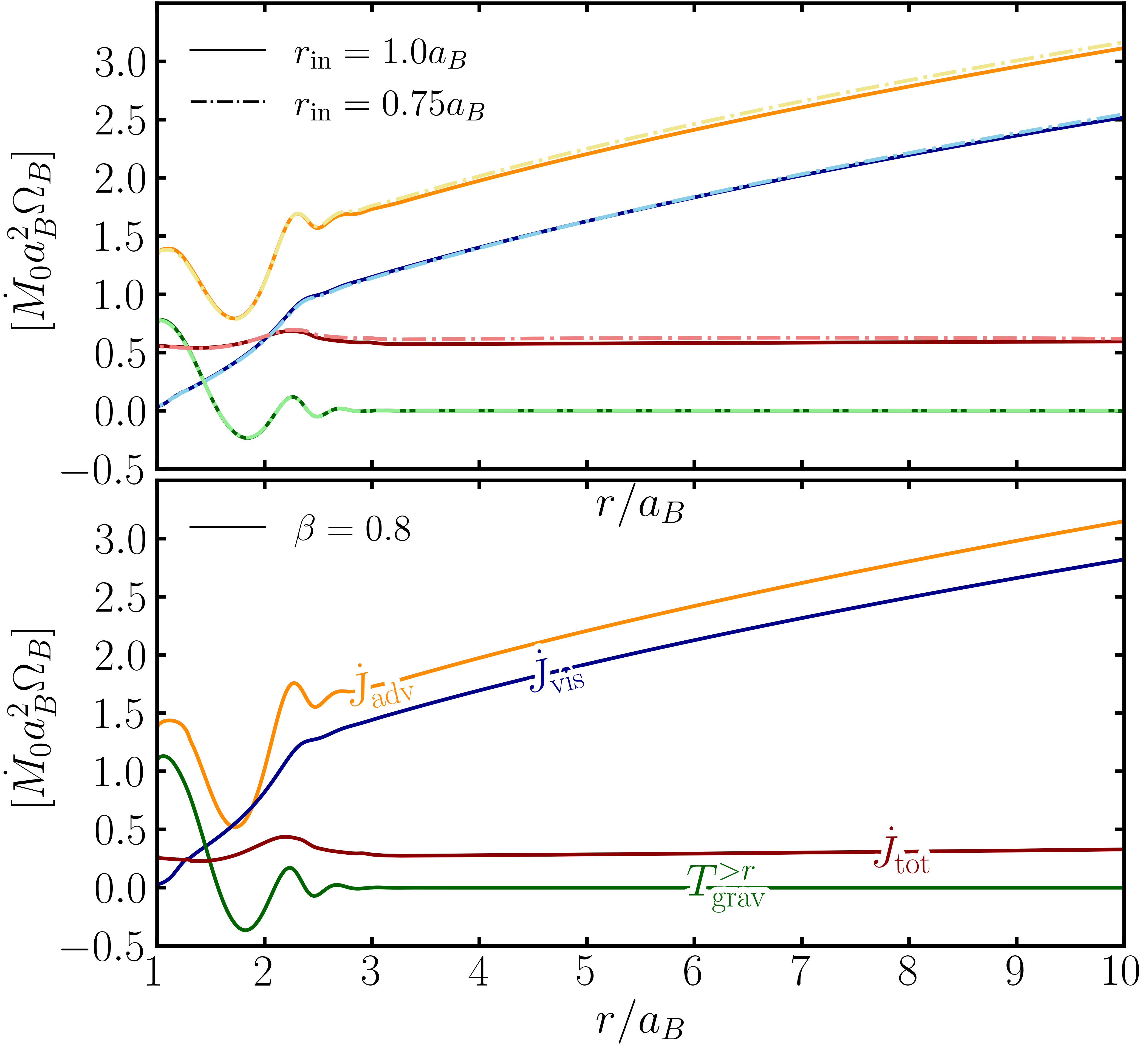}
\end{center}
\vspace{-0.3cm}
\caption{Radial profile of various torque components in simulation F1.0 and T with $\beta=1.0$ (upper panel) and B0.8 (lower panel). 
In the upper panel, the solid lines denote the results from Run F1.0, with inner boundary located at $a_{\rm{B}}$. The dashed lines are from simulation T, with $r_{\mathrm{in}}=0.75a_{\rm{B}}$. The torques from the two simulations overlap, indicating that the net angular momentum transfer rate is independent of the inner boundary location. 
In the lower panel, the total torque exerted on the binary is less than $(3/8)\dot{M}a_{\rm{B}}^2\Omega_{\rm{B}}$, showing binary inspiral. Different from the upper panel, the gravitational torque close to the inner boundary increases much faster with decreasing $r$. 
\label{fig:am-innertest} \label{fig:am-spiral}}
\vspace{-0.3cm}
\end{figure}



\begin{figure*}[htbp]
\begin{center}
\includegraphics[width=1.0\textwidth,trim={0cm 0cm 0cm .0cm},clip]{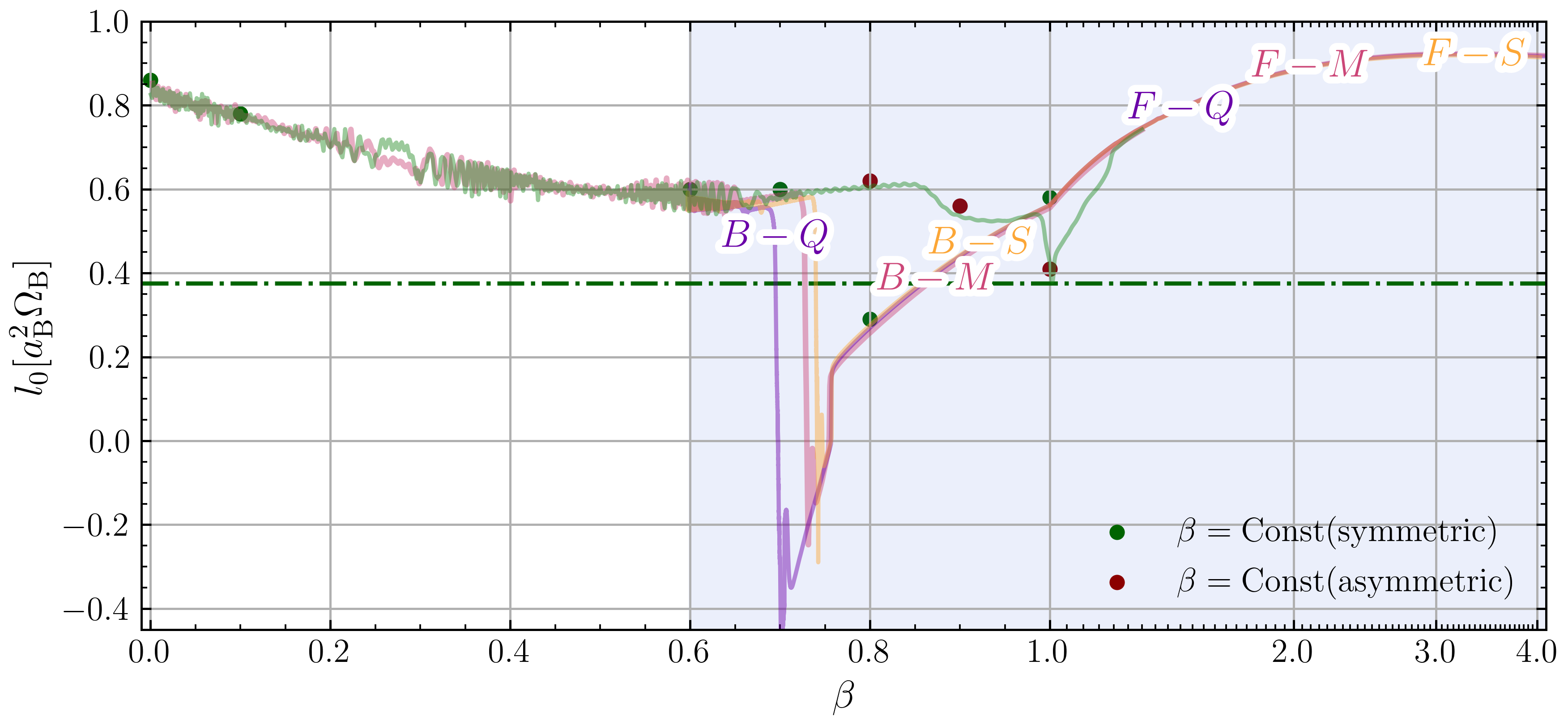}
\end{center}
\vspace{-0.3cm}
\caption{The dependence of specific angular momentum transfer from the disk to the binary on the cooling parameter $\beta$. For better illustration, we use linear scale along x axis when $\beta \in [0.0, 1.0]$ and log scale when $\beta \in [1.0,4.0]$. The green(red) line denotes simulation F(B), with $\beta$ increases from 0 to 1.3(decreases from 1.3 to 0). The dark red(dark green) dots are simulations with constant $\beta$ starting from the axisymmetric(non-axisymmetric) initial condition. In the blue shaded region, we carry out a series of long-term simulations F-Q, F-M, F-S, B-Q, B-M/B, B-S, changing the timescale that $\beta$ varies. Convergence is achieved in the F-series of runs at $\beta>1.0$, and in the B-series of runs as the timescale increases. It is clear that simulations B and F behave qualitatively similar, and the ``dip" is postponed (to large $\beta$) in simulation F with a smaller magnitude. The horizontal green dash-dotted line denotes the threshold for binary orbital expansion. Only in simulation B-series, a small window of binary orbital inspiral exists between $0.7 \lesssim \beta \lesssim 0.9$. \label{fig:am-sum}}
\vspace{-0.0cm}
\end{figure*}

To explore the dependence of the accretion variability on the cooling time ($t_{\mathrm{cool}}$ or $\beta$), we measure the accretion rate at the inner boundary, and the results are shown in Figure \ref{fig:acc-vari}. 
Corresponding to the two transitions in disk morphology(see Section \ref{sec:result-morphology}), there are also two evident changes in accretion variability as $\beta$ increases. 

First, in the locally isothermal limit, two accretion periodicities (at frequency of $2\Omega_{\rm{B}}$ and $\Omega_{\rm{B}}/5$) are most prominent, related to the accretion streams coupled to the binary and the high-density lump orbiting around the cavity, respectively \citep{2008ApJ...672...83M,2013MNRAS.436.2997D,2017MNRAS.466.1170M}. 
As $\beta$ increases to $0.7$, the lump is gradually replaced by a collection of spirals, which consequently suppresses the $\Omega_{\rm{B}}/5$ periodicity. 
As shown in the spacetime diagram of the accretion rate in Figure \ref{fig:acc-colormap}, the lump orbiting around the binary gradually disappears as {$\beta$ increases} (e.g. in the uppermost panel, the red region denoting the highest accretion rate traces the evolution of the lump: the radial movement of the lump can be clearly identified.)
After this first transition, the accretion rate is dominated by the narrow accretion streams which accumulate into spirals at the cavity wall. The natural frequency of this process is $\sim 2\Omega_{\rm{B}}$ (see Figure \ref{fig:morphology-halforbit}).

Second, further increase in $\beta$ leads to the total suppression of accretion variability (see the lowermost panel of Figure \ref{fig:acc-colormap} and the rightmost panel of Figure \ref{fig:acc-vari}), which happens at $0.7 \lesssim \beta \lesssim 0.8$. 
Materials are accreted following the two-armed spiral density waves coupled to the binary. 
Unlike the lump and accretion streams in the cases of small $\beta$, the tightly-wound density waves stay rather stable, which leads to no notable variability in accretion rate.

\subsection{Angular Momentum Transfer}
\label{section:angular-momentum-transfer}

\begin{figure*}[htpb]
\begin{center}
\includegraphics[width=0.8\textwidth,trim={0cm 0cm 0cm .0cm},clip]{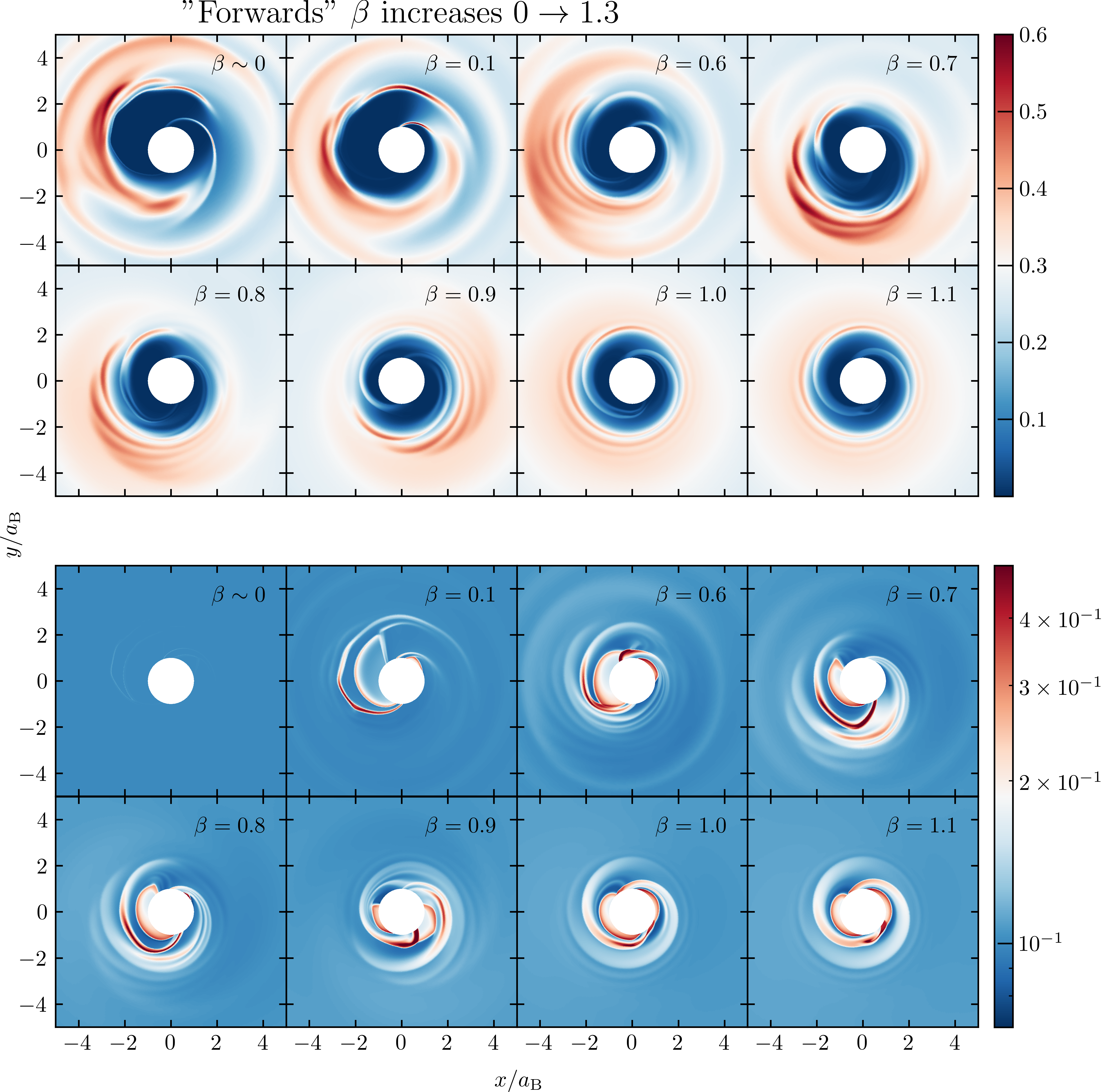}
\end{center}
\vspace{-0.3cm}
\caption{
Same as Figure \ref{fig:morphology-backward}, but for Run F, $\beta$ increases from 0 to 1.3. Similar to Figure \ref{fig:morphology-backward}, two transitions in disk morphology can be identified though $\beta$ corresponding to the transition become larger.
}\label{fig:morphology}
\end{figure*}

Figure \ref{fig:am-innertest} (upper panel) shows the profiles of the time-averaged advective, viscous, gravitational and total torque from Run F0 and T ($r_{\text{in}}=a_{\rm{B}}$ and $r_{\text{in}}=0.75a_{\rm{B}}$). These two simulations yield the same torque profiles, suggesting that the results are independent of inner boundary location.

The dependence of the net angular momentum transfer rate on $\beta$ is illustrated in Figure \ref{fig:am-sum}. 
The green and red solid lines are the results taken from the two fiducial long-term simulations F and B with continuous varying $\beta$. 
Also plotted in Figure \ref{fig:am-sum} are the constant $\beta$ runs for $\beta = {0.1, 0.6, 0.7, 0.8, 1.0}$ starting from an axisymmetric initial condition ({i.e. starting from the initial condition described in Section \ref{sec:setup}}), and $\beta = {0.8, 0.9, 1.0}$ restarted from the steady state of Run F0. 
The horizontal green dashed line denotes the threshold for binary orbital expansion.
For Run F (the solid green line), within the range of our parameter survey the binary would always outspiral, although the specific angular momentum transfering from the disk to the binary exhibits a ``v'' shape: it first decreases to a minimum value, and then increases as $\beta$ becomes larger. 

For Run B, in a small window of $0.7 \lesssim \beta \lesssim 0.9$, the binary shrinks. To examine this case in detail, we show the radial profile of various torques of Run B0.8 in Figure \ref{fig:am-spiral} (lower panel). It appears that the gravitational torque near the inner boundary is stronger, and the gap between viscous torque and advective torque is smaller in Run B0.8 as compared to the locally isothermal runs. 
In Figure \ref{fig:morphology-backward}, the two-armed trailing spirals also have the highest density when $\beta=0.8$, providing the strongest gravitational torque. To ensure the narrow window for binary inspiral is an actual physical result, we perform two additional simulations B-Q and B-S decreasing $\beta$ from 1.0 to 0.6 with different sweep speed $d (t_{\rm{interval}}/P_{\rm{B}})/ d (\beta/0.1) = 2000$ and $6000$. It can be seen from Figure \ref{fig:am-sum} that the $\beta$ value corresponding to the sharp transition of angular momentum transfer rate from outspiral to inspiral converges to $\beta \sim 0.75$. We also extend our parameter survey to $\beta = 4.0$ in Run F-Q, F-M, and F-S. These three runs converge nicely though having different sweep speed. It is noticeable that when $\beta \sim 1.2$, Run F also converges with the F-series simulations which start from $\beta=1.0$.
Though we validate the simulation results through varying the sweep speed, this window for binary inward migration is quite narrow and we postpone further investigation of binary's orbital evolution in the follow-up work. 




\subsection{``Hysteresis" in Disk Evolution}
\label{section:hysteresis}

{In the previous sections, we have focused on Run B (see Figure \ref{fig:morphology-backward}).
}
Since we do not have a clear physical reason for the second transition in disk morphology (at $\beta \sim 0.9$), we also run simulations with a different initial condition by restarting from the steady state of the locally isothermal run (referred to as the 'perturbed' initial condition for convenience), represented by Run F (see Figure \ref{fig:morphology}). While the results in disk morphology, accretion variability and angular momentum transfer agree for $\beta \lesssim 0.7$ and $\beta \gtrsim 1.2$, there are large differences between the two kinds of runs for intermediate $\beta$'s. 

As discussed in Section \ref{section:accretion-variability} and Section \ref{section:angular-momentum-transfer}, the transition from orbiting lump to ``twists" corresponds to the disappearance of $\Omega_{\rm{B}}/5$ accretion variability and decreases in the specific angular momentum flux from the disk to the binary. And the second transition to a circular cavity and tightly-wound two-armed spirals corresponds to the complete suppression of all accretion periodicity and increment in angular momentum flux. Comparing Run B (Figure \ref{fig:morphology-backward}) and Run F (Figure \ref{fig:morphology}), it is evident that the two transitions occur with smaller $\beta$ in the former. In Figure \ref{fig:am-sum}, Run B (represented by the green solid line) also shows a transition with smaller $\beta$ and a smaller minimum specific angular momentum. Although there is a small window of orbital decay for Run B, the binary always experiences orbital expansion for Run F. 

Our study shows that for some values of $\beta$, there may exist two steady states which can be reached, depending on the initial condition. The reason for this ``hysteresis'' is unclear, but it suggests the existence of two degenerate steady-state solutions for the disk. \citet{2021ApJ...914L..21D} also found hysteresis behaviour in their study on binary eccentricity. The hysteresis found in our work could be present for rapidly evolving binary systems. We plan to investigate this problem in our follow-up work, where we resolve the mini-disks and streams within $r < a_{\rm{B}}$. 

\section{Summary and Discussion} \label{sec:conclusion}

\subsection{Summary}

We have conducted a series of 2D hydrodynamical simulations of circumbinary disks, focusing on the effect of thermodynamics of the gas. We adopt the simple $\beta$-prescription for the gas cooling time, $t_{\text{cool}}=\beta/\Omega_{\rm{K}}$, and consider equal-mass, circular binaries {with circumbinary disks having aspect ratio $h=0.1$. We also set the kinematic viscosity according to the $\alpha$ prescription.} 
We assessed the inner disk structure, accretion variability, and orbital evolution of the central binary due to the presence of the circumbinary disk after the inner disk has reached the quasi-steady state. 

As the parameter $\beta$ increases from 0 (locally isothermal limit) to 4.0, two transitions exist in the disk morphology, accretion variability, and binary orbital evolution. These transitions happen at different $\beta$ values and may depend on our choice of the initial condition. Our main findings are as follows.

\quad (i) Disk morphology: As $\beta$ (or the cooling time) increases, the high-density lump orbiting at the edge of the cavity first changes to a collection of spirals that steadily rotate around the binary and  then changes to tightly-wound two-armed spirals coupled to the binary.

\quad (ii) Accretion variability: As the cooling time becomes longer, the accretion variability of the binary is gradually suppressed. The long-term variability at the frequency $\Omega_{\rm{B}}/5$ associated with the lump disappears when $\beta \sim 0.7-1.0$, but other variabilities remain. Further increase of $\beta$ leads to the complete suppression of the $2\Omega_{\rm{B}}$ variability, and binary accretes steadily.

\quad (iii) Binary's orbital evolution: The angular momentum transfer between the circumbinary disk and the binary is affected by the gas cooling. The specific angular momentum transferred from the disk to the binary exhibits a decrease then an increase with increasing cooling time. For most $\beta$-values, the binary orbit expands while accreting from the disk. The binary could shrink in a narrow window of $0.7 \lesssim \beta \lesssim 0.9$ if the disk starts from a symmetric initial condition.

\subsection{Limitations and Prospects}

As an initial study, our work has focused on circumbinary disks (CBDs) with aspect ratio $H/r=0.1$ and viscosity parameter $\alpha=0.1$, hosting equal-mass circular binaries. While the physics and cooling prescription included in our work is highly simplified, and the parameter range explored in our simulations is limited, the results highlight the importance of thermodynamics, and potentially radiation transport, in governing the dynamics and evolution of CBDs.

One major limitation of this work is that
we have excised the central circular region ($r<a_{\rm{B}}$) to maximize the computational efficiency, thus do not accommodate the accretion flows/streams around the binary and their mini-disks. As is demonstrated in Appendix \ref{appendix}, given the parameters we have (carefully) chosen, our results are largely unaffected by this treatment, 
and a quasi-steady state for accretion and angular momentum transfer can be achieved without resolving the mini-disks. On the other hand, we show in Appendix \ref{appendix} that this approach is likely inappropriate to study thinner disks ($H/r<0.1$), where the eccentricity of the cavity becomes significantly larger and an excised central domain would erroneously prevent materials on eccentric orbits around the cavity from returning to the disk \citep{2022ApJ...932...24T}.

In the near future, we will improve upon this work in several aspects. First, we will fill the ``hole" at the center and resolve the mini-disks around
individual binary components. This would open up the wide parameter space that is left unexplored in this work. 
Second, three-dimensional hydrodynamic simulations \citep{2019ApJ...875...66M} and MHD simulations to capture the more realistic MRI (e.g.  \citealt{2012ApJ...749..118S,2012ApJ...755...51N,2021ApJ...922..175N}) would be another natural extension in the future.
Finally, treating thermodynamics of the disk by thermal relaxation towards equilibrium profile, as presented in this work, is highly simplified. Radiative cooling \citep{2019A&A...627A..91K} and turbulent heating should be incorporated in future studies. 

We thank the anonymous referee for the insightful comments and suggestions, which increases the scope of this work. We also thank Zoltan Haiman, Haifeng Yang, Douglas N. C. Lin, Yixian Chen, and Xiaochen Sun for useful discussions, as well as the KITP program "Bridging the Gap: Accretion and Orbital Evolution in Stellar and Black Hole Binaries" for helpful communications. This work is supported by the Dushi fund at Tsinghua University, and in part by the National Science Foundation under Grant No. NSF PHY-1748958. Numerical simulations are conducted in the Orion cluster at Department of Astronomy, Tsinghua University, and in TianHe-1 (A) at National Supercomputer Center in Tianjin, China.


\appendix

\section{The Validity of Simulations Excising the Central Domain}
\label{appendix}
\begin{figure}[htbp]
\begin{center}
\includegraphics[width=0.4\textwidth,trim={0cm 0cm 0cm .0cm},clip]{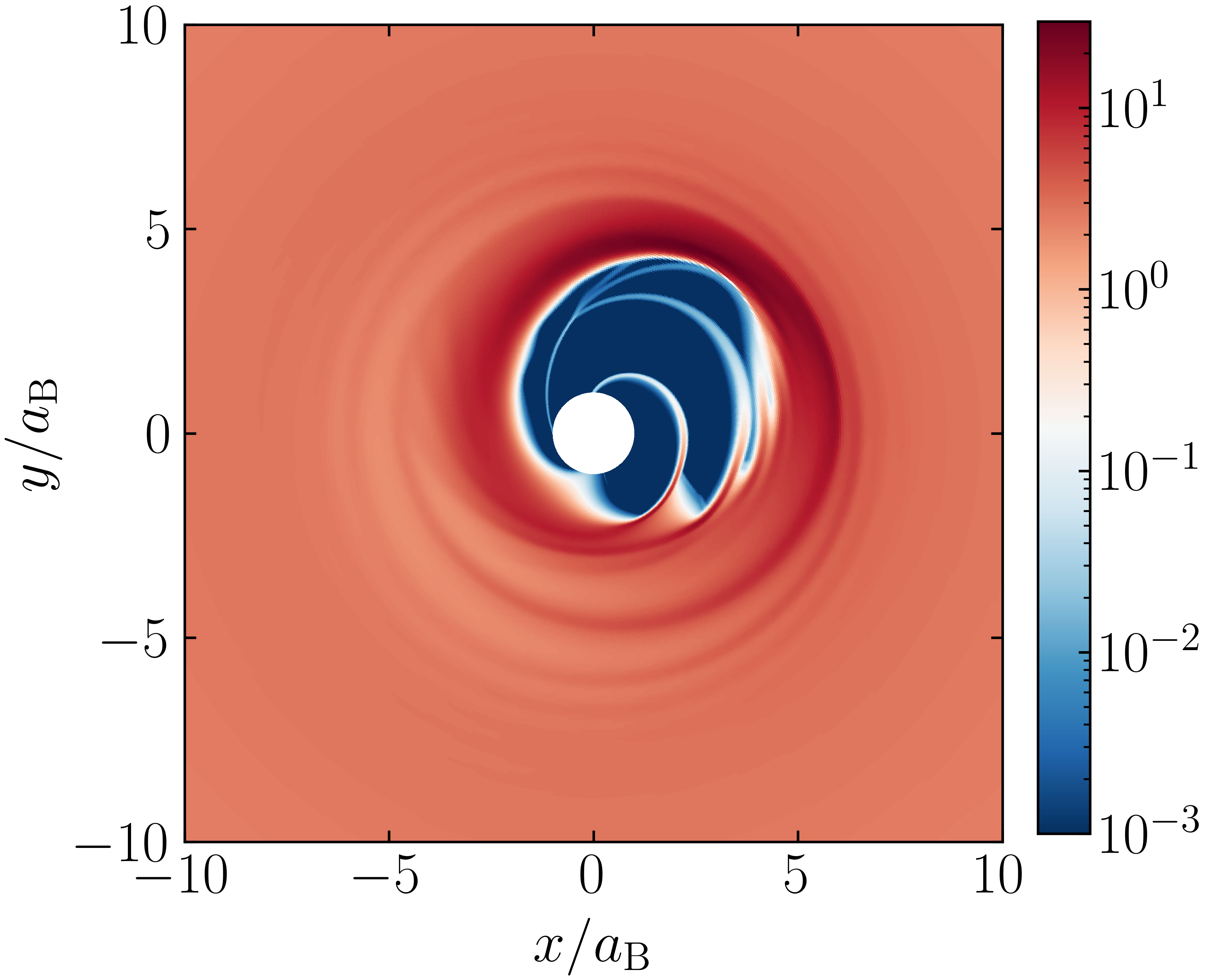}
\end{center}
\vspace{-0.3cm}
\caption{Snapshots of disk surface density from thin disk simulation ($h=0.03$). The scale of the colormap is logrithmic to incorporate the large density variation. \label{fig:morphology-appendix}}
\vspace{-0.3cm}
\end{figure}

\begin{figure}[htbp]
\begin{center}
\includegraphics[width=0.48\textwidth,trim={0cm 0cm 0cm .0cm},clip]{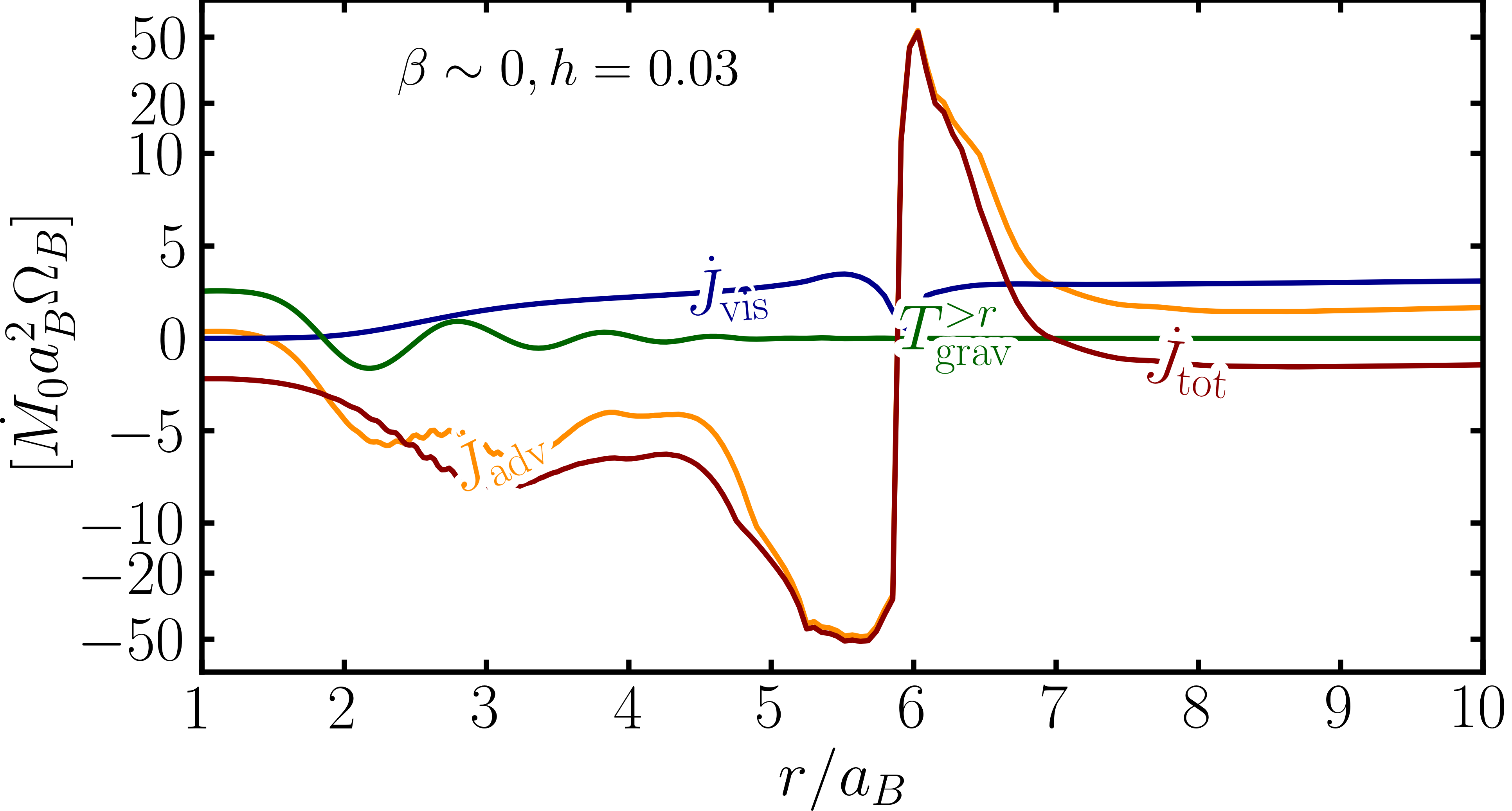}
\end{center}
\vspace{-0.3cm}
\caption{Radial profile of various torque components in thin-disk simulation ($h=0.03$). To incorporate the large variations, the scale in the vertical direction is linear within $[-10,10]$ and logrithmic for larger or smaller values. Even being averaged over 10,000 orbits, the total torque exhibit extremely large variation, showing that the inner CBD did not reach a quasi-steady state. \label{fig:am-appendix}}
\vspace{-0.3cm}
\end{figure}

In this Appendix, we discuss the validity of simulations excising the central domain by considering disks with smaller aspect ratio, which is better applicable to CBDs around supermassive black hole binaries. In doing so, 
we perform a series of locally isothermal simulations having the aspect ratio of $h=0.03$ for 50,000 orbits. All the parameters expect the aspect ratio are the same as our fiducial Run T0. Figure \ref{fig:morphology-appendix} shows a simulation snapshot ($t=50,000 P_{\rm{B}}$) and Figure \ref{fig:am-appendix} shows the radial profile of various torque components averaged over 10,000 orbits. 

Excising the central domain could induce problems when studying the binary orbital evolution with the disk aspect ratio smaller than $0.1$. As found by previous studies (e.g. \citealt{2022MNRAS.513.6158D}), CBDs having a smaller aspect ratio possess a more eccentric cavity, making the periapse of the cavity become closer to the inner boundary. For such an eccentric disk, the ``diode'' inner boundary condition is no longer valid. Materials pass close to the binary will slingshot to a larger radius \citep{2022ApJ...932...24T} (i.e. we expect both inflow and outflow at $r_{\mathrm{in}}=a_{\rm{B}}$), while the ``diode'' inner boundary forbids such materials to return to the simulation domain.
As can be seen in Figure \ref{fig:am-appendix}, even averaged over $10^4$ binary orbits, the total angular momentum flux fails to approach a steady state, and exhibits highly significant radial variations. It is also clear that the radial variation is primarily in the form of the advective flux, consistent with the expectation of missing returning materials from the inner boundary. The radial location where this variation is the most significant likely corresponds to the location where the highly eccentric materials are originated.
We have also tried a series of simulations with $h=(0.05,0.08,0.09)$, and the variations in $\dot J_{\mathrm{tot}}$ decrease with increasing aspect ratio. The torque variations are within $10\%$ only when the aspect ratio increases till $h=0.1$, thus justifying the simulations presented in this paper. Considering that eccentric binary will slingshot gas from $r<a_{\rm{B}}$ to outer region, we conclude that only equal mass circular binary with aspect ratio $h\geq0.1$ ensures the correctness of simulations when excising the central domain. 
Due to the same reason, we also anticipate that the exclusion of inflow will influence the disk morphology, thus caution must be exercised when interpreting simulation results of certain previous works (e.g.  \citealt{2022A&A...660A.101P,2022A&A...664A.157S}).
CBDs with smaller aspect ratio will be thoroughly studied in our follow-up study after resolving the central domain.

\bibliography{CBD-simulation}{}
\bibliographystyle{CBD-simulation}

\end{document}